\pdfoutput=1
\documentclass[english,aps,prl,twocolumn, superscriptaddress, amsmath]{revtex4-1}

\usepackage[T1]{fontenc}
\usepackage[latin9]{inputenc}
\usepackage{color}
\usepackage{units}
\usepackage{amssymb}
\usepackage{graphicx}
\usepackage{esint}
\usepackage{bm}
\usepackage{natbib}
\usepackage[colorlinks=true, citecolor=red]{hyperref}
\setcitestyle{journalcolor= blue}

\usepackage{graphicx}
\usepackage{dcolumn}
\usepackage{bm}

\usepackage{babel}

\begin{document}

\title{Effect of  dimensionality on the vortex-dynamics in type-II superconductor}

\author{Hemanta Kumar Kundu}
\affiliation{Department of Physics, Indian Institute of Science, Bangalore 560012, India}
\author{Kazi Rafsanjani Amin}
\affiliation{University Grenoble Alpes, CNRS, Grenoble INP, Institut N\'{e}el, 38000 Grenoble, France}
\author{John Jesudasan}
\affiliation{Tata Institute of Fundamental Research, Mumbai 400005, India}
\author{Pratap Raychaudhuri}
\affiliation{Tata Institute of Fundamental Research, Mumbai 400005, India}
\author{Subroto Mukerjee}
\affiliation{Department of Physics, Indian Institute of Science, Bangalore 560012, India}
\author{Aveek Bid}
\email{aveek@iisc.ac.in}
\affiliation{Department of Physics, Indian Institute of Science, Bangalore 560012, India}

\begin{abstract}

We explore the effects of sample dimensionality on vortex pinning in a type-II, low-$T_C$, s-wave superconductor, NbN, in the presence of a perpendicular magnetic field, $H$. We find significant differences in the phase diagrams in the magnetic field--temperature plane between  3-dimensional (3D) and 2-dimensional (2D) NbN films. The differences are most striking close to the normal-superconductor phase transition. We establish that these variances have their origin in the differing pinning properties in two different dimensions. We obtain the pinning strength quantitatively in both the dimensions from two independent transport measurements performed in two different regimes of vortex-motion -- (i) thermally assisted flux-flow (TAFF) regime and (ii) flux flow (FF) regime. Both the measurements consistently show that both the pinning potential and the  zero-field free-energy barrier to depinning in the 3D superconductor are at least an order of magnitude stronger than that in the 2D superconductor. Further, we probed the dynamics of pinning in both 2D and 3D superconductor through voltage fluctuation spectroscopy. We find that the mechanism of vortex pinning-depinning is qualitatively similar for the 3D and 2D superconductors. The voltage-fluctuations arising from vortex-motion are found to be correlated only in the 2D superconductor. We establish this to be due to the presence of long-range phase fluctuations near the Berezinskii-Kosterlitz-Thouless (BKT) type superconducting transition in 2-dimensional superconductors. 
 
\end{abstract}

\maketitle

 Beyond the lower critical field ($H_{C1}$), magnetic field penetrates a type-II superconductors (SC) in the form of topological defects known as vortices or fluxoids~\cite{tinkham2004introduction}. A vortex comprises of  a circulating supercurrent  and encloses a magnetic flux  quantized to $\phi_0 = h/(2e)$. The interaction between the vortices is repulsive, producing a periodic structure called the Abrikosov vortex lattice (VL)~\cite{ABRIKOSOV1957199,levy2013magnetism}. The periodicity of this lattice gets distorted in the presence of inhomogeneities and fluctuations (both thermal and quantum)  having energy-scales comparable to the elastic-energy of the VL~\cite{doi:10.1080/000187300412257,HIGGINS1996232,PhysRevB.52.1242,PhysRevB.55.6577,ganguli2015disordering,PhysRevLett.67.3444,de2007density}. For a disordered superconductor, the phase-diagram is complex.  Depending on the relative strengths of the pinning-potential, thermal energy and elastic energy of the VL, a gamut of phases like vortex-solid, vortex-fluid and vortex-glass can exist~\cite{HIGGINS1996232,PhysRevB.43.130,RevModPhys.66.1125}. Several different types of phase-transitions/cross-overs connecting these phases have been predicted (and in some cases experimentally verified) in a three-dimensional superconductor (3D-SC)~\cite{RevModPhys.66.1125,PhysRevLett.122.047001,PhysRevB.95.134505,PhysRevLett.63.1511,PhysRevLett.76.2555}.  The phase-diagram becomes even more interesting in a two-dimensional superconductor (2D-SC) in which even for arbitrarily small pinning-strengths, the long-range translational order of the VL is lost although rotational order survives~\cite{PhysRevB.43.130,RevModPhys.66.1125,PhysRevLett.41.121,Larkin1979,tsen2016nature,PhysRevLett.70.670,PhysRevLett.65.923,benyamini2019absence}.  Thus an understanding of the interplay of fluctuations, disorder, and dimensionality, is of paramount importance in describing the dynamics and the related phase transitions of the vortex state of a type-II SC. 

In an idealized SC devoid of defects, the vortices are free, and an infinitesimal current or thermal excitation is enough to cause them to move leading to dissipation~\cite{PhysRev.140.A1197}. Thus, in such a system, in the presence of a magnetic field, the true zero-resistance state can survive only at zero-temperature. Vortex pinning is essential to restore the zero resistance state in a disordered SC.  Hence, an understanding of pinning mechanism, methods to controllably create defects with high pinning potential or to produce commensurable pinning effects using artificially-created ordered-series of defects have always been at the forefront of fundamental as well as applied research~\cite{Villegas1188,PhysRevB.77.060506,PhysRevB.81.092505,PhysRevLett.92.180602,PhysRevLett.98.117005,PhysRevLett.78.2648}.

Among the many exciting features of high-T$_C$ SC, pinning of vortices in the mixed-state has always attracted much interest~\cite{RevModPhys.59.1001, FEIGELMAN1990177,PhysRevLett.65.259,WORTHINGTON1990417,PhysRevLett.64.966,PhysRevLett.63.1511,PhysRevB.66.024523,PhysRevB.64.184523}. Even after decades of research, its exact origin and consequences are not well understood~\cite{RevModPhys.66.1125}. This is partly due to the complications present in high-T$_C$ materials due to substantial thermal fluctuations or the `irreversibility-line' in the $H-T$ phase diagram~\cite{PhysRevLett.60.2202}. To circumvent these impediments, we probed the vortex-lattice, both in 2D- and 3D- limits in a conventional type-II superconductor. Specifically, we looked at thin films of NbN which are known to be in the strong-pinning limit~\cite{PhysRevB.79.094509,PhysRevB.83.214517,PhysRevB.96.054509}. The aim was to isolate and investigate only the dimensional effects on vortex dynamics. Local probes like scanning tunneling microscopy (STM) or magnetic force microscopy (MFM) correlate the static position of the vortices with atomic-scale structural defects~\cite{Auslaender2008,Pardo1998,Hoffman1148,RevModPhys.79.353}. Magneto-transport measurements, on the other hand, probe the variation in the global dynamics of vortices with pinning potential strength or with dimensionality~\cite{RevModPhys.66.1125,PhysRevLett.70.670}. These two complementary techniques together provide a detailed picture of local pinning forces as well as the collective dynamics of vortices. In a series of previous publications, some of us  looked in detail, using low-temperature STM,   at the local-dynamics of vortices in SC~\cite{PhysRevLett.107.217003, PhysRevB.96.054509,PhysRevB.79.094509,PhysRevB.85.014508,PhysRevLett.111.197001}.  In this article, we look at the temporal and spatial correlations of the vortex-dynamics through detailed magneto-transport measurements.

Bulk NbN is a well known, s-type, conventional type-II superconductor well described by  Bardeen-Cooper-Schrieffer (BCS) theory~\cite{PhysRevLett.106.047001,PhysRevB.79.094509}. High-quality NbN superconducting films of different thickness and desired disorder levels can be grown with excellent control~\cite{PhysRevB.91.054514,PhysRevB.85.014508}. This makes NbN an ideal system to compare and contrast superconductivity in two different dimensions -- 3D and 2D. The superconducting coherence length obtained from critical field measurement is $\sim$ 6~nm for NbN; any NbN film of thickness lesser than 6~nm behaves as a 2D-superconductor~\cite{PhysRevLett.111.197001}. 

We studied the superconductor-normal phase diagram in the perpendicular magnetic field-temperature ($H$-$T$) plane for NbN films of two thicknesses -- 68~nm (3D-SC) and 3~nm (2D-SC).  The films were patterned into four-probe configurations, with four 10~nm/60~nm Cr/Au electrical contacts, each  2~mm wide, 1~mm in length and separated from each other by 200~$\mu$m thermally deposited on them through a metal-mask. The measurements were done in a pumped 2~K cryostat (equipped with an 8~T superconducting-magnet) with the films immersed in the $^4$He-exchange gas to ensure good thermalization. Special care was taken to thermalize and low-pass filter the measurement wires (with a cut-off frequency of 500~KHz) going into the cryostat. 

The resistive transitions are found to be strikingly different in these two cases. We find that in 2D-SC, close to the Berezinsky-Kosterlitz-Thouless (BKT) transition~\cite{KT, berezinski1973jetp} temperature  $T_{BKT}$, a very small magnetic field is enough to give rise to dissipative transport even in the limit of vanishingly-small current. In contrast, for the 3D-SC, a  finite magnetic field is required to get dissipative transport over the entire phase-space. In both cases, we have defined the critical-temperature to be the $T$ at which the sample resistance $R$ becomes 1\% of its normal state resistance $R_N$. To understand this significant difference in the response of superconductivity in different dimensions to a perpendicular magnetic field ($H$), we analyzed the pinning properties of vortices of both 2D- and 3D-SC. We conclude from two independent sets of measurements -- (a) temperature dependence of the magnetoresistance and (b) $H$ dependence of the critical current density $J_C$ -- that pinning of vortices is more than one order of magnitude stronger in 3D-SC than in 2D-SC. To understand the dynamical process of pinning of vortices, we looked into voltage fluctuations in these films as a function of $H$ at temperatures close to $T_C$. We find that the voltage fluctuations are an order of magnitude slower in 2D-SC as compared to the 3D-SC. We also find that the dependence of the relative variance of voltage fluctuations on $H$ is the same in both cases showing that pinning-depinning of vortices is the dominant source of noise in both 2D- and 3D-SC. Computations of higher moments of fluctuations indicate the presence of strong correlations between the vortex motion in 2D while it is essentially uncorrelated in the 3D-SC. 

\begin{figure}
	\begin{center}
		\includegraphics[width=0.48\textwidth]{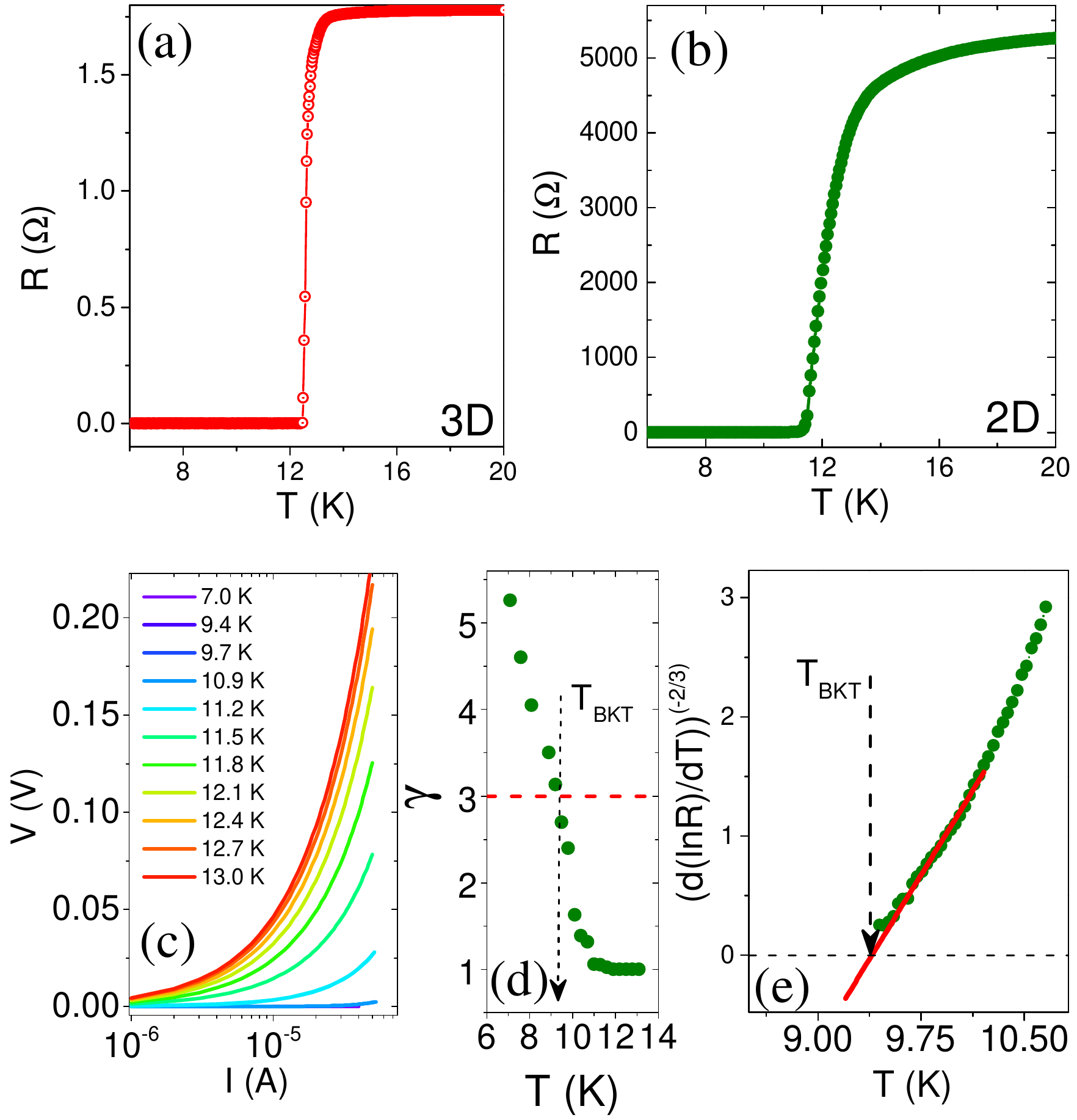}
		\small{\caption{Resistance versus temperature plot for (a) 68~nm and (b) 3~nm thick NbN film. (c) Plots of the non-linear current-voltage characteristics of the 3~nm NbN film at several temperatures around $T_{BKT}$. (d) Plot of $\gamma$ versus temperature. $\gamma(T_{BKT}) = 3$ is marked by red dashed line, this yield $T_{BKT}=9.4$~K. (e) Fit of the measured temperature dependence of resistance to Eqn.~\ref{bkt_RTeq}. The green filled circles are the data and the solid red line is the fit. $T_{BKT}$ extracted from the intercept is found to be $9.4~K$.
				\label{fig:RT}}}
	\end{center}
\end{figure}

Fig.~\ref{fig:RT}(a) shows the temperature dependence of the resistance of the 3D-SC NbN film. The mean field transition temperature $T_C$ for this film is 12.4~K.  The corresponding data for the 3~nm film are plotted in Fig.~\ref{fig:RT}(b). Our previous studies of the temperature dependence of super-fluid density have established that 3~nm thick NbN film undergoes a BKT transition (which is a hallmark of 2D-SC) while the 68~nm film is a BCS 3D-SC~\cite{PhysRevB.91.054514,PhysRevLett.111.197001,PhysRevLett.106.047001}. In this paper, we employ a different approach and identify $T_{BKT}$ for the 3~nm NbN superconducting film from electrical transport measurements. One can identify $T_{BKT}$ by two different electrical transport measurements. The first comes from the measurements of current--voltage ($I$--$V$) characteristics in the superconducting regime. According to Ginzburg-Landau Coulomb gas description of 2D-SC, a finite driving electrical current flowing through the sample leads to proliferation of free vortices from dissociation of bound votex-antivortex pairs. These freely-flowing vortices cause phase-slips giving rise to dissipation in the system which follows, in the low-current range, a non-linear $I$--$V$ relation: $V \sim I^\gamma$~\cite{RevModPhys.59.1001,PhysRevB.39.9708,PhysRevB.42.2242,PhysRevB.94.085104,Reyren1196,tsen2016nature}. In this prescription, $T_{BKT}$ is identified by the criterion $\gamma(T_{BKT}) = 3$.  The  $I$-$V$ characteristics measured at different $T$ are plotted in fig.~\ref{fig:RT}(c). The corresponding $\gamma$ values are plotted in fig.~\ref{fig:RT}(d). From this plot, we identify $T_{BKT}$  to be $\sim$9.4~K.

\begin{figure}[t]
	\begin{center}
		\includegraphics[width=0.48\textwidth]{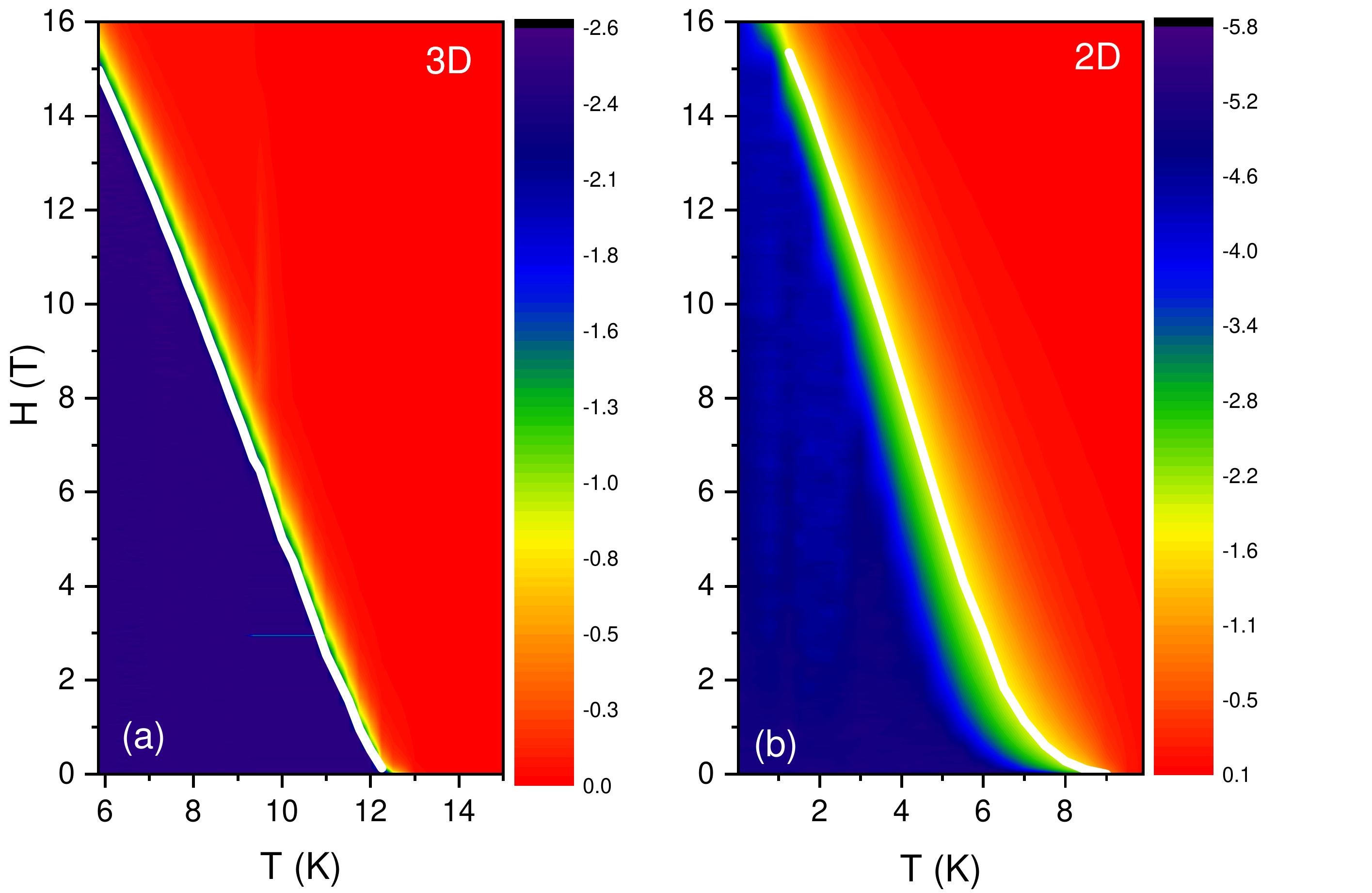}
		\small{\caption{(a) Plot of  the logarithm of the normalized resistance, log$(R/R_N)$ in  the $H$--$T$ plane for the 3D-SC. (b) Corresponding plot for the 2D-SC. The solid white line in both the plots show the $H_{C}$ [defined as $R(T,H=H_C)= 0.01R_N$] as a function of $T$.
				\label{fig:HT}}}
	\end{center}
\end{figure}

The second method to obtain $T_{BKT}$ is from the temperature dependence of resistance $R$ of the SC. Near $T_{BKT}$, for a 2D-SC, it is known to follow the relation:
\begin{equation}
R = R_0 \exp\bigg(\frac{b_R}{(T-T_{BKT})^{1/2}}\bigg)
\label{bkt_RTeq}
\end{equation}
where $b_R$ is a measure of  the strength of interaction between the vortices and anti-vortices~\cite{RevModPhys.59.1001,PhysRevLett.55.2156,PhysRevB.21.1806}. In this temperature regime vortex anti-vortex pairs unbind thermally. Their proliferation leads to phase-fluctuations and consequently to the suppression of superconductivity. To estimate $T_{BKT}$, we fit the $R-T$ data to equation~\ref{bkt_RTeq} as shown in fig~\ref{fig:RT}(e). This procedure yields $T_{BKT}\sim 9.4$~K which is in agreement with $T_{BKT}$ extracted from non-linear $I$-$V$ characteristics. This value of $T_{BKT}$ also matches closely with that obtained from measurements of super-fluid number-density by us~\cite{PhysRevLett.107.217003,PhysRevLett.111.197001}. Note that this method is an approximation and is not compelling enough to establish the two-dimensional nature of superconductivity. The important drawbacks of this method are -- (i) finite size effects are ignored; and (ii) it is valid over a very narrow temperature range $T_{BKT}\leq T<T_C$. Nonetheless, it is a handy technique to estimate the BKT transition temperature in materials where the 2D nature of SC is already established. 

\begin{figure}[t]
	\begin{center}
		\includegraphics[width=0.48\textwidth]{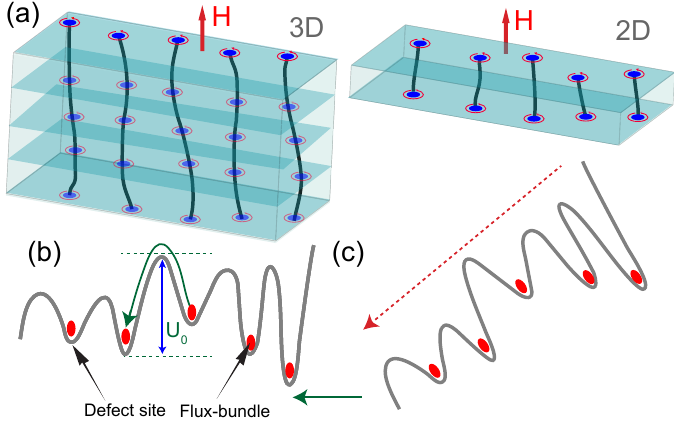}
		\small{\caption{(a) Schematic showing the difference in pinning strengths between 3D-SC and 2D-SC (b) Schematic illustrating the concept of vortices getting pinned at local pinning-potential minima at $J\ll J_C$. The free-energy barrier to hop from one pinning-site to another is $U_0(H)$. The vortices can only hop from one pinning-site to another by thermal activation -- this is the thermally-activated flux flow (TAFF) regime. (c) In the presence of  a driving current $J \geq J_C$ through the SC, the Lorentz-force  lowers the hopping-barrier by effectively tilting the potential landscape. When $\textbf{F}_\textbf{L}(H) \geq \textbf{F}_\textbf{P}(H)$, the vortices can flow freely giving rise to an increase in dissipation -- this is the flux flow (FF) regime.
				\label{fig:scheme}}}
	\end{center}
\end{figure}

We now turn to the response of the superconducting state to a magnetic field $H$ applied perpendicular to the plane of the film. We obtain the phase diagram in the $H$--$T$ plane from magnetoresistance measurements performed at different temperatures. The phase diagrams are shown in fig.~\ref{fig:HT}(a) and (b) for 3D-SC and the 2D-SC respectively. There is a significant difference in the way dissipation arises close to transition temperature in these two cases. For example at $T = 0.85$~$T_C$, for 3D-SC  4.12~T is needed to get finite dissipation. On the other hand, an order of magnitude smaller field -- $\sim$0.44~T -- is enough to induce dissipative transport in the 2D-SC at  $T=0.85$~$T_C$. A possible reason for this fragility of SC in 2D as compared to its 3D counterpart can be the difference in pinning-strength of vortices in the two cases. In ultra-thin 2D films, the vortices are effectively pancake-like as opposed to the tube-like structure in 3D. Consequently, in 3D-SC the vortices naturally get pinned at several different pinning sites along the thickness of the film. The pancake vortices in 2D-SC do not get pinned as much and hence can move more easily, giving rise to dissipation. The idea is illustrated  in Fig.~\ref{fig:scheme}(a)  which represents schematically the  pinning in 3D-SC and 2D-SC.

\begin{figure}[t]
	\begin{center}
		\includegraphics[width=0.48\textwidth]{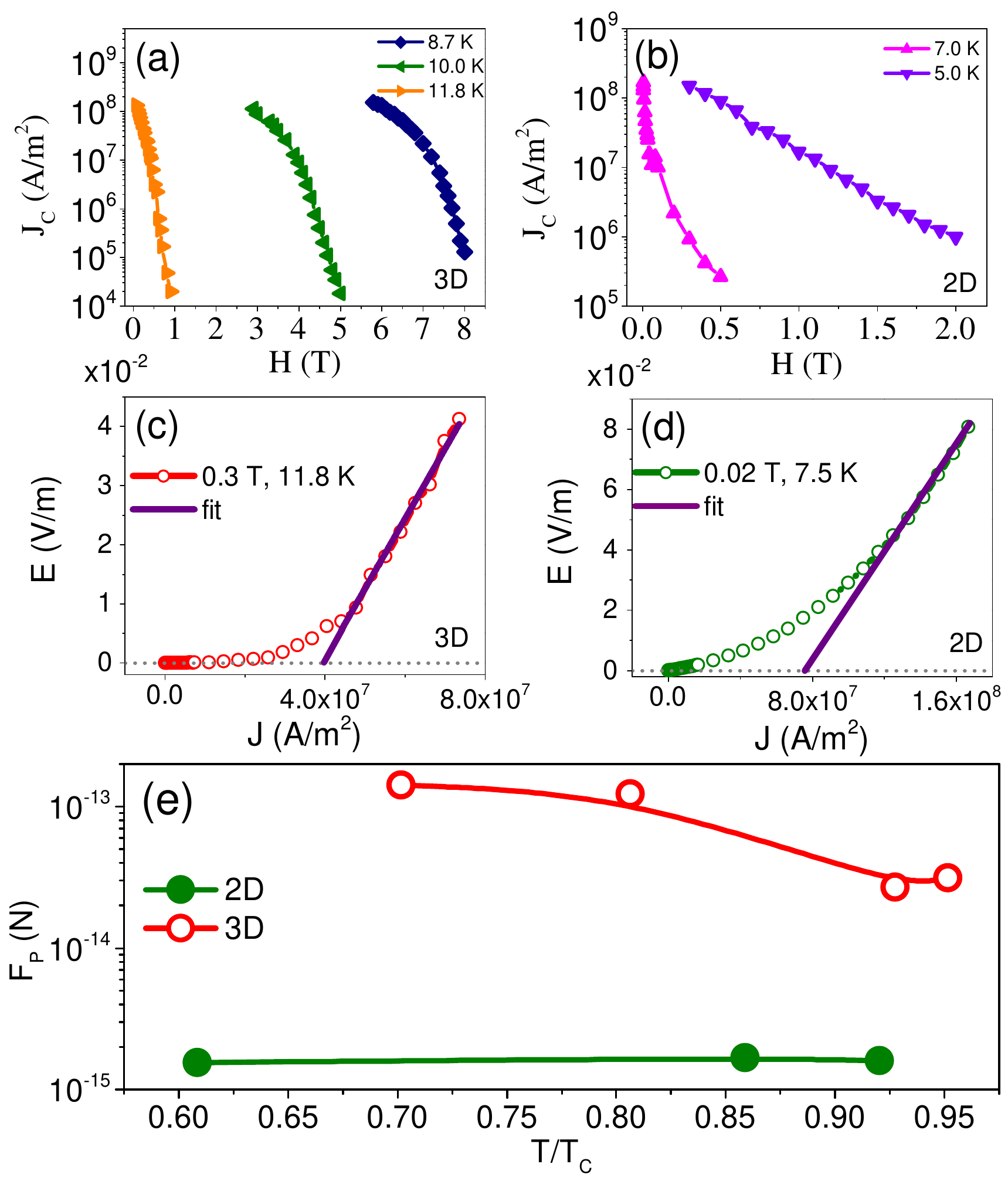}
		\small{\caption{(a) Plots showing the variation of critical current density $J_C$ versus the magnetic  field  $H$ at few selected temperatures for the 3D-SC. (b) Plot of the corresponding data for the 2D-SC (c) Plot of $E$ versus $J$ for the 3D-SC measured at $T/T_C$=0.95 and $H$=0.3~T. (d) Plot of $E$ versus $J$ for the 2D-SC measured at  $T/T_{BKT}$=0.9 and $H$=0.02~T. The intercept of the solid purple lines in both (c) and (d) give the value of corresponding $J_C$. (e) Plot of pinning force per vortex $F_P$ versus $T/T_C$. One can see that the $F_P$ in 3D-SC is at least an order of magnitude stronger than in the 2D-SC.
				\label{fig:Fp}}}
	\end{center}
\end{figure}

To test this hypothesis of differing pinning strengths in 2D-SC and 3D-SC,  we estimated the pinning force and pinning energy from two independent measurements in both 2D and 3D superconductors. Transport measurements evaluate the average pinning force on vortices considering the pinning of the flux-bundles due to sample inhomogeneities and also the interaction between the vortices. For values of magnetic fields much smaller than the upper-critical field, $H<<H_{C2}$, the vortices are separated by large distances $\sim\sqrt{H/\phi_0}$ as compared to the penetration depth, $\lambda$. In this limit, the vortices can be considered as independent, non-interacting objects. Recall that a current density $\textbf{J(H)}$ flowing through the SC perpendicular to the magnetic field $\textbf{H}$ leads to a Lorentz force $\mathrm{\textbf{F}_\textbf{L}(H) = d_{film}\textbf{J}\times \hat{\textbf{n}}\phi_0}$ acting on each vortex-line. Here $\hat{\textbf{n}}$ is the unit-vector parallel to $\textbf{H}$, $\mathrm{d_{film}}$ is the film-thickness and $\phi_0=h/(2e)$ is the flux-quantum which is the net flux threading a single vortex.  This force tends to aid the flux-bundles overcome the free-energy barriers related to the pinning effect of inhomogeneities in the sample and move them in a direction perpendicular to both $\textbf{J}$ and $\textbf{H}$.   It is opposed by the average pining-force per vortex, ${\textbf{F}_\textbf{P}(\mathrm{H})}$, leading to the net force on the vortex being $\mathrm{\textbf{F}_{\textbf{net}}(H) = \textbf{F}_\textbf{L}(H)+\textbf{F}_\textbf{P}(H) = d_{film}\textbf{J}(H)\times \hat{\textbf{n}}\phi+\textbf{F}_\textbf{P}(H)}$. When the Lorentz force overcomes the pinning force, the vortices get depinned and begin to flow. This motion induces an electric potential in a direction perpendicular to both $\textbf{H}$ and $\textbf{F}_\textbf{L}$, i.e. in a direction parallel to $\textbf{J}$.  Thus, work is done, and energy is dissipated -- the vortex motion leads to a finite resistance. The critical-current density $\textbf{J}_\textbf{C}$ is defined as the maximum current density that can flow through the SC  without de-pinning the vortices, or equivalently $\lim_{H \to 0}\textbf{F}_{\textbf{net}}(H) = 0$.  This force-balance equation relates the maximum magnitude of the pinning-force to that of the critical-current density:

\begin{equation}
\mathrm{F_P(H) = d_{film} J_C \phi_0}
\label{pinningForce}
\end{equation}
\begin{figure}[t]
	\begin{center}
		\includegraphics[width=0.48\textwidth]{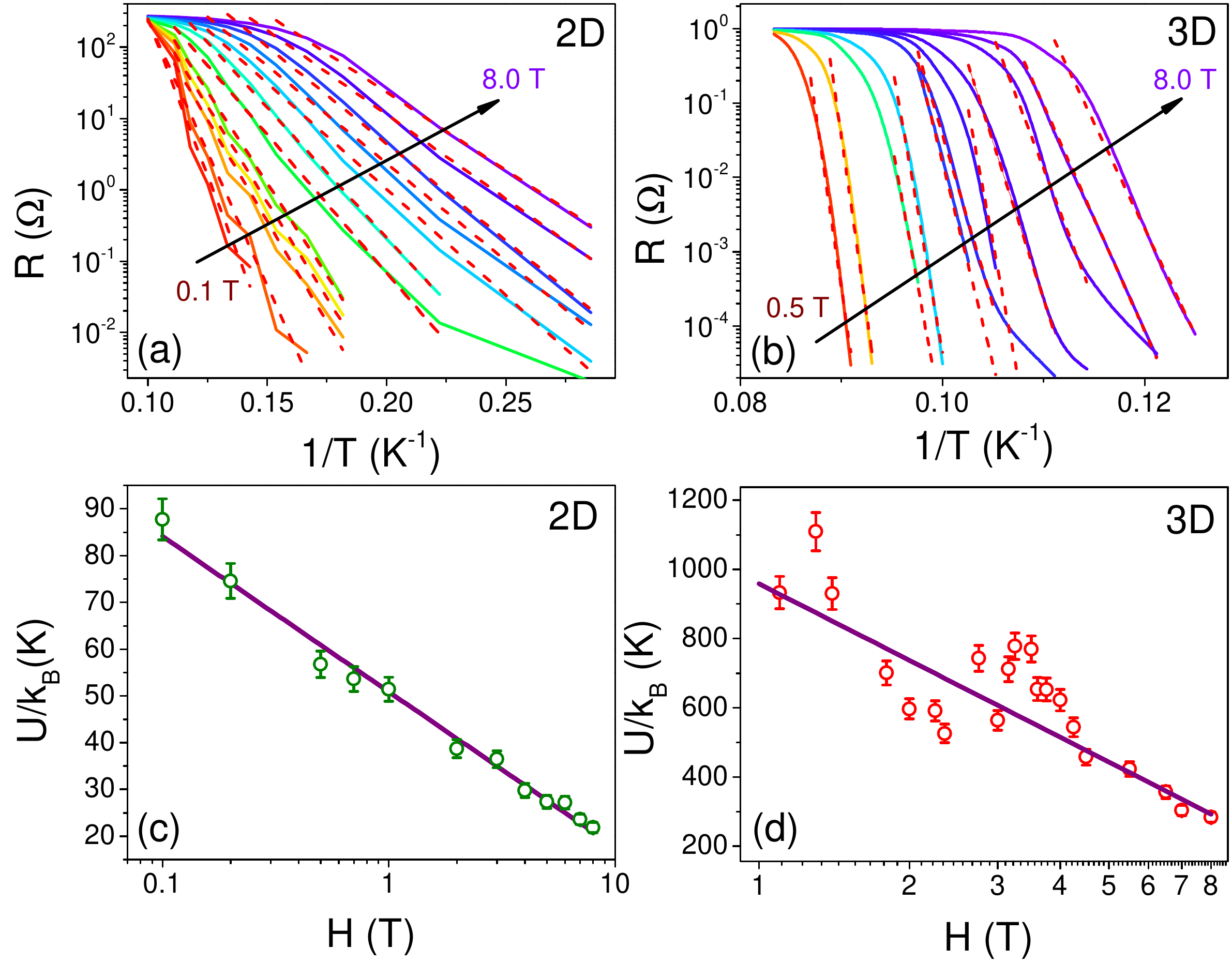}
		\small{\caption{(a) and (b) Plots of R, in log scale, versus $1/T$  at different magnetic fields for the 2D-SC and 3D-SC respectively.  The red dashed  lines in both the plots are the Arrehenius fits. The measurements were all performed in the TAFF regime. (c) The green open circles are the extracted activation energies $U(H)$ as a function of $H$ for the 2D-SC. The purple line is the fit to the relation $U(H) = U_0\ln(H_0/H)$, yielding $U_0/k_B$=14.5~K.  (d) $U(H)$ for the 3D-SC has a non-monotonic dependence on $H$ for the 3D-SC. The purple lines is a plot of the relation $U(H) = U_0\ln(H_0/H)$ with $U_0/k_B$=280~K.
				\label{fig:UB}}}
	\end{center}
\end{figure}

\noindent The limit $H\rightarrow 0$ ensures that the interactions between the vortices are irrelevant.  

We obtained $J_C(H)$ as a function of $H$ from measurements of $J$ versus $E$ at different $T$ and $H$.  $J_C$ was defined operationally as the value of current-density at which the potential drop across the film equaled 1~mV/m.  The measured variation of $J_C$ with $H$ is shown in Fig.~\ref{fig:Fp}(a) and Fig.~\ref{fig:Fp}(b) for the 3D- and 2D-SC respectively. 

The value of $J_C(H)$ were also extracted using an alternate independent method, by defining it to be the value of $J$ at which the linear-fit to the $E$-$J$ plots intersects the $J$-axis, here $E$ is the electric field in the plane of the film (see Fig.~\ref{fig:Fp}(c) and (d) for representative plots). The validity of this method stems from the fact that above $J>J_C$, flux-flow (FF) regime sets in and resistance in this regime, $R_{FF}(H)$  is independent of the current; hence the $E$-$J$ curve is linear. These two independent methods give values of $J_C(H)$ which match within a factor of 2-3.

	The average pinning force on each vortex-line $F_P(H)$ was calculated using Eqn.~\ref{pinningForce}; the results are plotted in Fig.~\ref{fig:Fp}(e). It is observed that vortex pinning strength for the 3D-SC is more than an order of magnitude larger than that for the 2D-SC. This is in agreement with our observation of initiation of dissipation in 2D-SC at much smaller values of $H$ as compared to the 3D-SC. 

The pinning strength can also be determined from the temperature $T$ dependence of the resistance of SC at $J \ll J_C$ in the presence of $H$. At such low current densities, $F_L \ll F_P$ and flux-motion can take place only through thermal-activation over the free-energy barriers related to the pinning induced by inhomogeneities in the sample. This regime is known as thermally activated flux flow (TAFF).  The idea is illustrated in Fig.~\ref{fig:scheme}(b). The differential resistance in this regime, $R_{TAFF}$ has a thermally activated behavior: $\lim_{J\to 0} R_{TAFF}(T,H) = R_0\exp(-U(H)/k_BT)$, where $U(H)$ is the energy-scale of the zero-field free-energy barrier to depinning~\cite{PhysRevLett.61.1662,PhysRevLett.67.1354,PhysRevLett.70.670}.  Figures~\ref{fig:UB} (a) and (b) show plots of $R$ in log-scale as a function of $1/T$ at different $H$ for 2D- and 3D-SC  respectively; the slopes of these plots give $U(H)$. The red dashed lines are the linear fits to the data. Note that in our measurements, this activated behavior is observed over four decades in resistance. Figure~\ref{fig:UB}(c) shows a plot of  of $U(H)$  versus $H$ for the 2D-SC in a semi-log scale;  the data were found to follow a logarithmic relation $U(H) = U_0\ln(H_0/H)$ with $U_0/k_B\sim 14.5K$. Theory predicts this logarithmic dependence of $U(H)$ on $H$ in the TAFF region  with $U_0 = \frac{\phi_0^2d}{256\pi^3\lambda^2}$~\cite{PhysRevLett.67.1354,PhysRevLett.70.670}. For NbN, $\lambda \approx 500$~nm giving  estimated $U_0/k_B$ to be $\sim$20~K for 3~nm thick NbN SC which agrees very well with our experimentally obtained number.

\begin{figure}[!]
	\begin{center}		\includegraphics[width=0.48\textwidth]{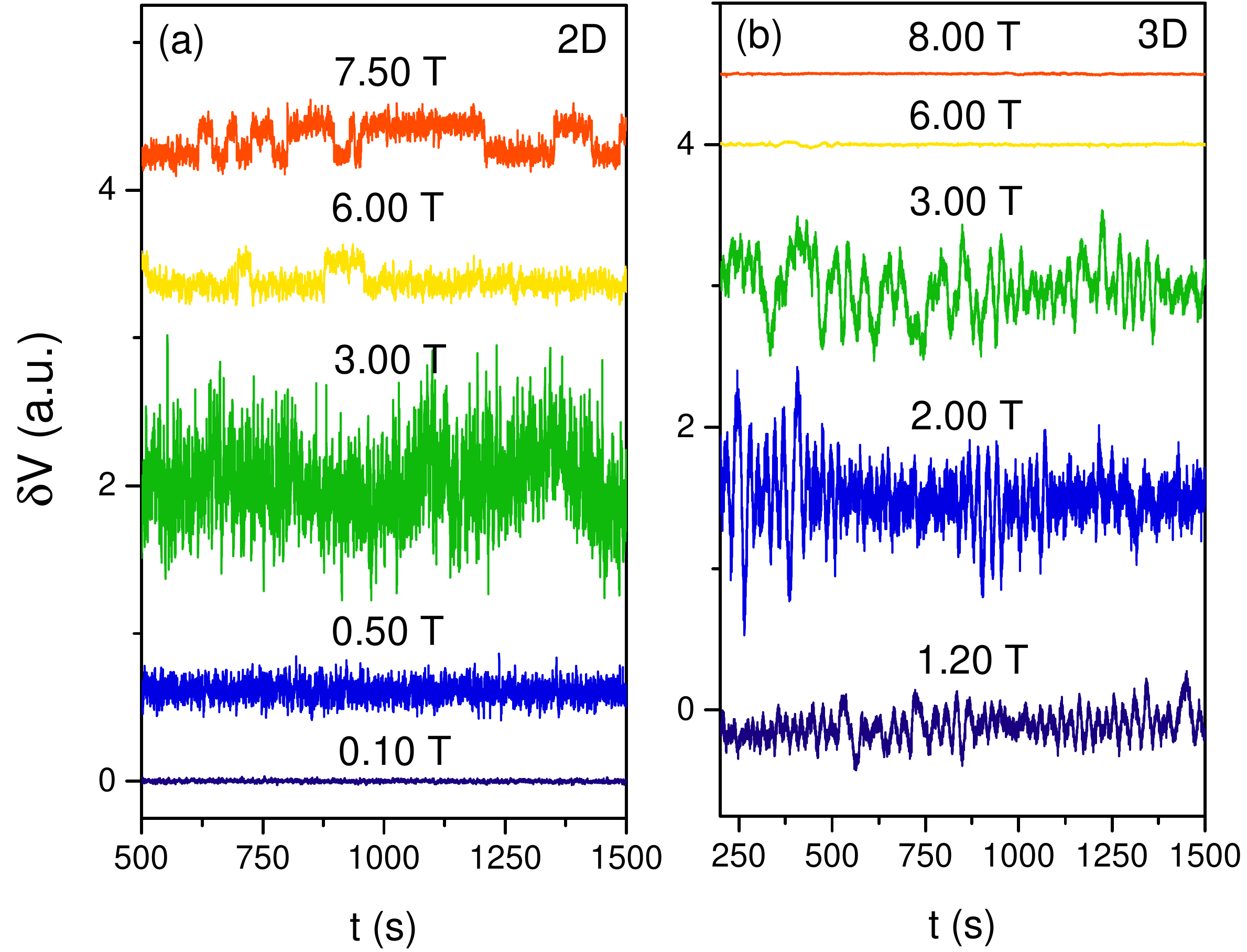}
		\small{\caption{(a) and (b) plots of  the time-series of voltage fluctuations at selected values of the magnetic field $H$ for the 2D- and 3D-SC respectively. The data have been verically shifted for clarity.						\label{fig:ts}}}
	\end{center}
\end{figure}

Figure~\ref{fig:UB}(d) shows that, unlike the case of 2D-SC, $U(H)$ for the 3D-SC varies with $H$ in a non-monotonic fashion.   The purple line is a plot of  $U(H) = U_0\ln(H_0/H)$ with $U_0/k_B$=280~K. The line matches the general trend of the plot of  $U(H)/k_B$ versus $H$ showing that  $U_0$ is approximately an order of magnitude larger in 3D-SC than its 2D counterpart.  The non-monotonic dependence of  $U(H)$ on $H$ has been observed previously in a similar superconducting system, thin-films of $\mathrm{Mo_xGe_{1-x}}$~\cite{PhysRevLett.70.670}. To the best of our knowledge,  there is no well-accepted explanation for this phenomenon. Upon comparison, we find that the activation energy scale $U(H)$ extracted from $R$ - $T$ plot is $\sim$~10 times larger for the 3D-SC as compared to the 2D-SC.  This observation is consistent with our estimate of comparative pinning forces in the two dimensions (Fig.\ref{fig:Fp}(e)). Using these two quantities ($F_P(H)$ and $U(H)$), we can estimate the spatial range of the pinning-potential (which one can heuristically  equate to the separation between pinning sites), to be $\sim$100~nm for both 2D and 3D-SC which roughly equals the impurity distribution in our films.

Thermally activated pining--depinning of vortices from sample inhomogeneities is a stochastic, dynamic process. As vortex motion gives rise to resistance in SC, a careful study of the dynamics of resistance-fluctuations {or equivalently, voltage fluctuations when the sample is driven by a with constant current} should provide insights into vortex-fluctuations. This motivated us to study voltage fluctuation and its higher-order moments  in both 2D-SC and 3D-SC at a fixed temperature $T = 0.95~T_C$ for the 3D-SC ($T = 0.97~T_{BKT}$ for the 2D-SC)  and at various values of the perpendicular magnetic field $H$.  

\begin{figure}[t]
	\begin{center}
		\includegraphics[width=0.48\textwidth]{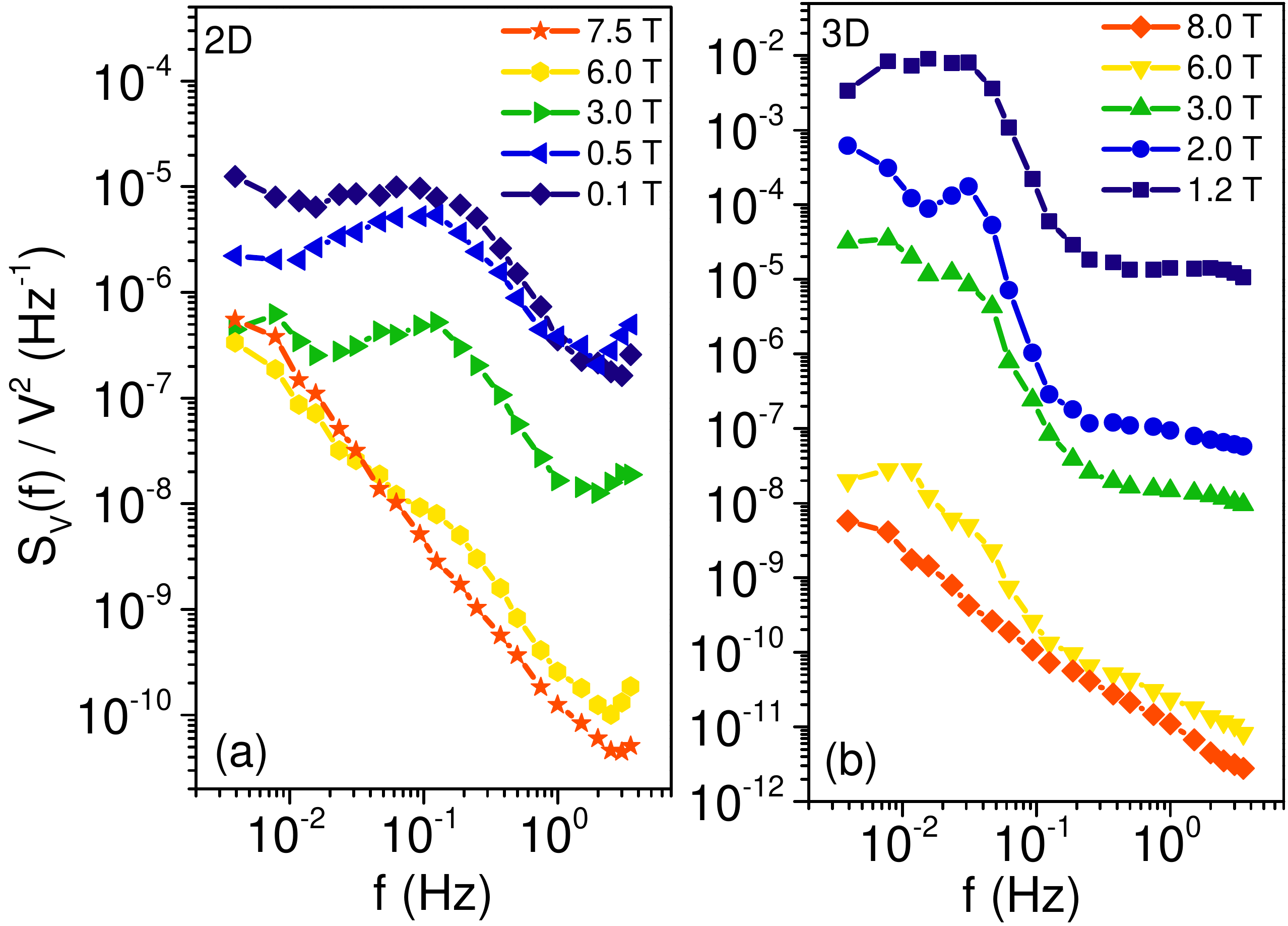}
		\small{\caption{(a) and (b) plots of  the normalized power spectral density of voltage fluctuations at a few representative magnetic fields $H$ in 2D- and 3D-SC respectively. 	The data  shown are at the same values of $H$ as in  Fig.~\ref{fig:ts}.			\label{fig:Svf}}}
	\end{center}
\end{figure}
To probe the voltage-fluctuations  (noise) and its statistics, we used a digital signal processing (DSP) based technique~\cite{scofield1987ac,Aveekthesis, ghosh2004set}. The technique allows us to measure the background noise as well as the dynamical noise from the sample at the same time. The film was biased by an ac current-density $J \ll J_C$ 

Plots of the voltage-fluctuation time-series for the 2D-SC and 3D-SC at a few selected values of $H$ are shown in fig.~\ref{fig:ts}(a) and (b). The corresponding normalized PSD of voltage fluctuations, $S_V(f)/V^2$ are plotted in in fig.~\ref{fig:Svf}(a) and (b) respectively. For fields $H>H_{C2}$, the dependence of $S_V(f)$ on $f$, in both 2D- and 3D-SC,  was found to be $S_V(f) \propto 1/f^\alpha $, where $\alpha \sim 1.1$. On the other hand, for  $H<H_{C2}$,  $S_V(f)$  was found to have developed a significant hump at a specific frequency $f_C$ riding on top of the $1/f^\alpha $ noise; $f_C$ for the 2D-SC and the 3D-SC were $\sim$ 0.125~Hz and 0.012~Hz respectively. This significant difference in the characteristic frequency scale of voltage-fluctuations in 2D-SC and 3D-SC  is a consequence of the fact that  the pinning-potential in the 2D-SC is significantly stronger than that in 3D.

\begin{figure}[t]
	\begin{center}
		\includegraphics[width=0.48\textwidth]{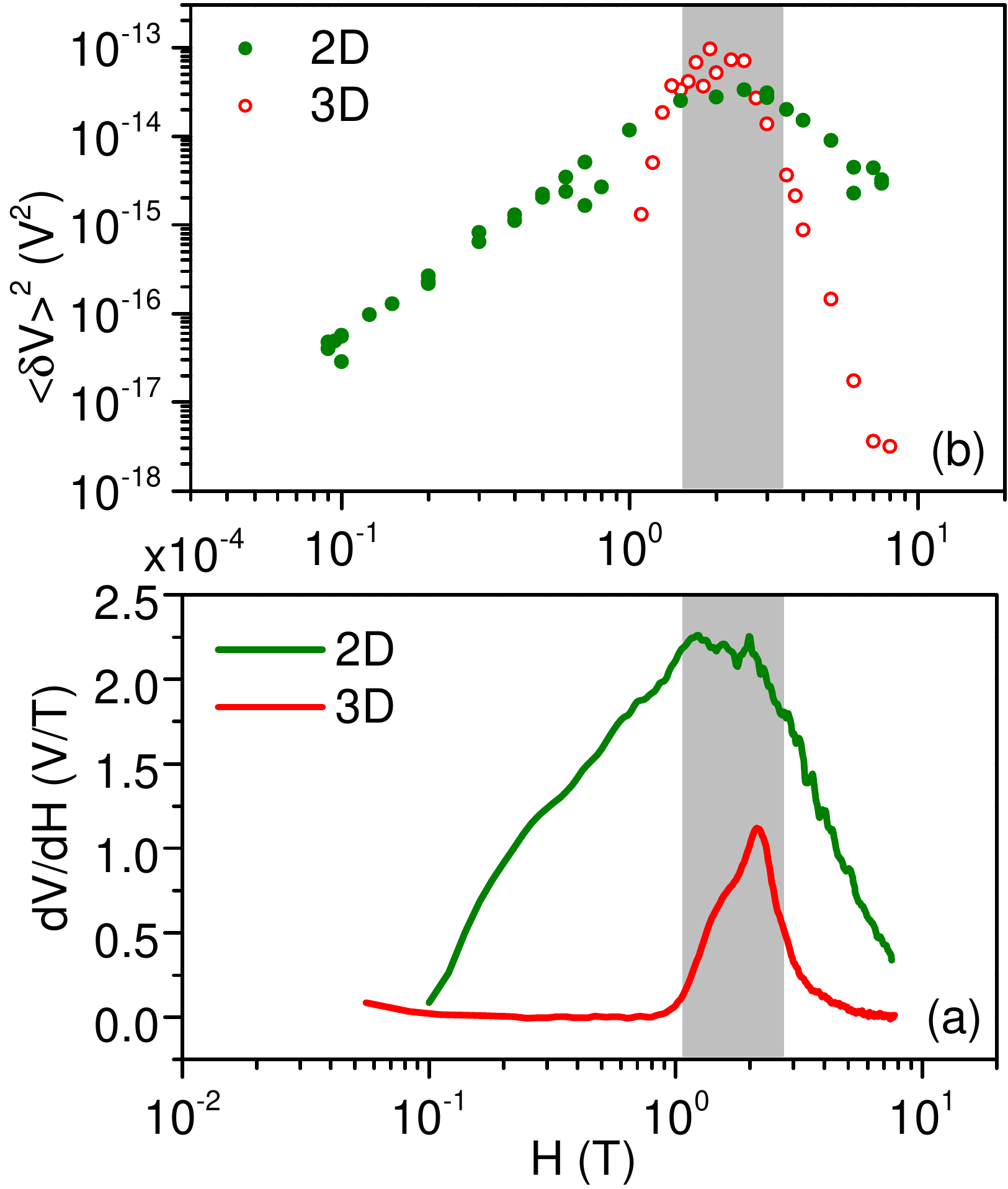}
		\small{\caption{(a) variance of voltage fluctuations $\mathcal{V}_{var}$ versus applied perpendicular  magnetic field $H$ for 2D-SC (green filled circles) and 3D-SC (red open circles). (b) Plot of $dV/dH$  versus $H$ for 2D-SC (dotted green line) and 3D-SC (solid red line). The gray shaded area indicates the range of  the magnetic field where the noise peaks for both 2D-SC and 3D-SC.
				\label{fig:variance}}}
	\end{center}
\end{figure}

The PSD of voltage fluctuation can be  integrated over the bandwidth of measurement to obtain the variance of voltage fluctuations $\mathcal{V}_{var}$~\cite{Aveekthesis, ghosh2004set}: 
\begin{eqnarray}
\mathcal{V}_{var}\equiv{\langle\delta V^2\rangle}=\int^{4 Hz}_{0.004 Hz}{S_V(f)df}.
\end{eqnarray}
\noindent Figure~\ref{fig:variance}(a) show plots of  $\mathcal{V}_{var}$ as a function of $H$.  In both 2D- and 3D-SC, we observe that the variance peaks around the same value of the magnetic field, $H\sim$~2~T. A probable origin of this noise can be fluctuations in $H$ as $dV/dH$ is quite sharp in this regime (Fig.\ref{fig:variance}(b)).  We rule  out this trivial explanation by noting that the  maximum voltage fluctuations arising from  fluctuations  in the magnetic field, $\delta H$ can be $\mathcal{V}_{var}^{mag}(H) =[(dV/dH)\delta H]^2 $. In our system the maximum value of $\delta H$ was measured to be $\mathrm{9\times 10^{-5}}$~T.   Using this value, we estimate the noise due to $H$-field fluctuations at $H$=2~T to be $\sim10^{-16}$~V$^2$ for the 2D-SC; this value is at least two-orders of magnitude smaller than the measured  $\mathcal{V}_{var}$ at 2~T.  

Instead, we propose that the observed noise arises from dynamic trapping-detrapping of vortices. The peak in the variance of voltage fluctuations for both 2D-SC and 3D-SC NbN films appear at the same magnetic field indicating that the magnitude of voltage-fluctuations depends only on the number of vortices - and not on the dimensionality of the system. Recall that the current density used during the noise measurements are minimal ($J\ll J_C$) and so the transport is by thermal-activation of vortices.  
Consider then the following scenario:  at low $H$, there are very few vortices. Consequently, both the voltage drop $V$ (caused by phase-slips due to vortex-drift under the driving current, hence proportional to the number of free vortices) and the voltage-fluctuations $\mathcal{V}_{var}$ (caused by fluctuation in the number of such  phase-slip events, i.e. proportional to the variance in the number-density of free vortices) are small. As $H$  increases, both  $V$  and $\mathcal{V}_{var}$  increase as  there are more vortices available to take part in these processes.  Beyond a certain field, almost all the pinning sites get accommodated with vortices, and there are not enough free \textcolor{red}{(empty)} pinning sites for the vortices to hop to; the fluctuations in the number of free-vortices and consequently the voltage-fluctuations decrease. 
\begin{figure}[t]
	\begin{center}
		\includegraphics[width=0.48\textwidth]{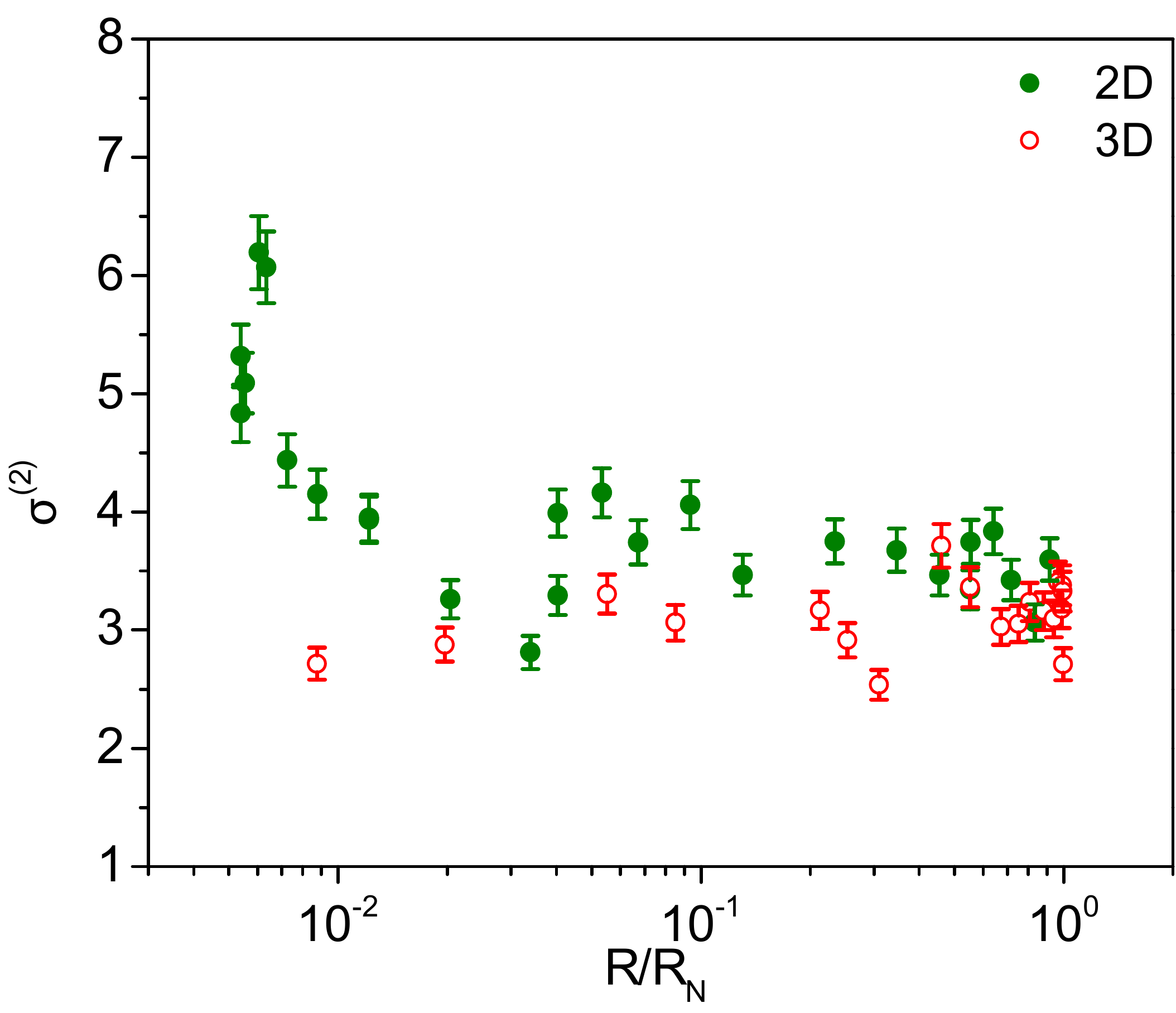}
		\small{\caption{Plots of normalized second spectrum ($\sigma^{(2)}$) versus applied perpendicular  magnetic field $H$ for 2D-SC (green filled circles) and 3D-SC (red open circles).  In 2D-SC, $\sigma^{(2)}$ $\sim$~3 at high field before it monotonically rises to about 7 near the SC-normal transition. On the other hand,  for the 3D-SC, $\sigma^{(2)}$ always fluctuates around 3, as for any Gaussian process. The sharp rise in $\sigma^{(2)}$, in the 2D case, appears at a  resistance  ($R/R_N \leq 0.01$) which is significantly less than the value  ($R/R_N\sim0.5$) at which the variance $\mathcal{V}_{var}$ peaks (\ref{fig:variance}(a)).
				\label{fig:SS}}}
	\end{center}
\end{figure}
The appearance of a  non-Gaussian distribution of the fluctuations in a system indicates correlated dynamics of the fluctuators (in our case, caused by stochastic pinning-depinning of vortices) in the system~\cite{PhysRevB.53.9753}.  As the correlation-length increases and eventually diverges, (as it happens close to criticality),  the fluctuations develop a prominent non-Gaussian component.   `Second-spectrum', which is the four-point correlation function of voltage-fluctuations calculated over a frequency-octave $(f_l,f_h)$, is extremely sensitive to the presence of non-Gaussian component (NGC) in voltage-fluctuations. Mathematically, it is given by:
\begin{eqnarray}
S_V^{f_1}(f_2)=\int_0^\infty \langle\delta V^2(t)\rangle\langle\delta V^2(t+\tau)\rangle cos(2\pi f_2\tau)d\tau
\label{Eqn:2spectrum}
\end{eqnarray}
Here $f_1$ is the center-frequency of the chosen octave in which $S_V^{f_1}(f_2)$ is calculated  and $f_2$ is spectral frequency. For any Gaussian random processes, the spectral response of $S_V^{f_1}(f_2)$ is independent of $f_2$. 

To calculate the second spectrum, we first make repeated measurement of $S_V(f)$ over a frequency-octave and obtain a `time-series' of $\mathcal{V}_{var}(t)$. The frequency-octave  is selected such that the sample noise is at least an order of magnitude higher than the background noise in order to avoid contamination of the data by the Gaussian background noise; in our case it was chosen to be $0.187$~Hz--$0.375$~Hz.  The PSD of this time-series of $\mathcal{V}_{var}(t)$ is the second-spectrum. A convenient representation of  the second spectrum is by normalizing it as follows:
\begin{eqnarray}
\sigma^{(2)}=\int_0^{f_h-f_l}S_V^{f_1}(f_2)df_2/\bigg[\int_{f_l}^{f_h}S_V(f)df\bigg]^2
\end{eqnarray}
For any Gaussian fluctuation process, $\sigma^{(2)}=3$;  any significant deviation from this value indicates the presence of long-range correlations between the fluctuators~\cite{PhysRevLett.111.197001,PhysRevB.94.085104,PhysRevB.85.045127,PhysRevB.90.085116}.

\begin{figure}[t]
	\begin{center}
		\includegraphics[width=0.48\textwidth]{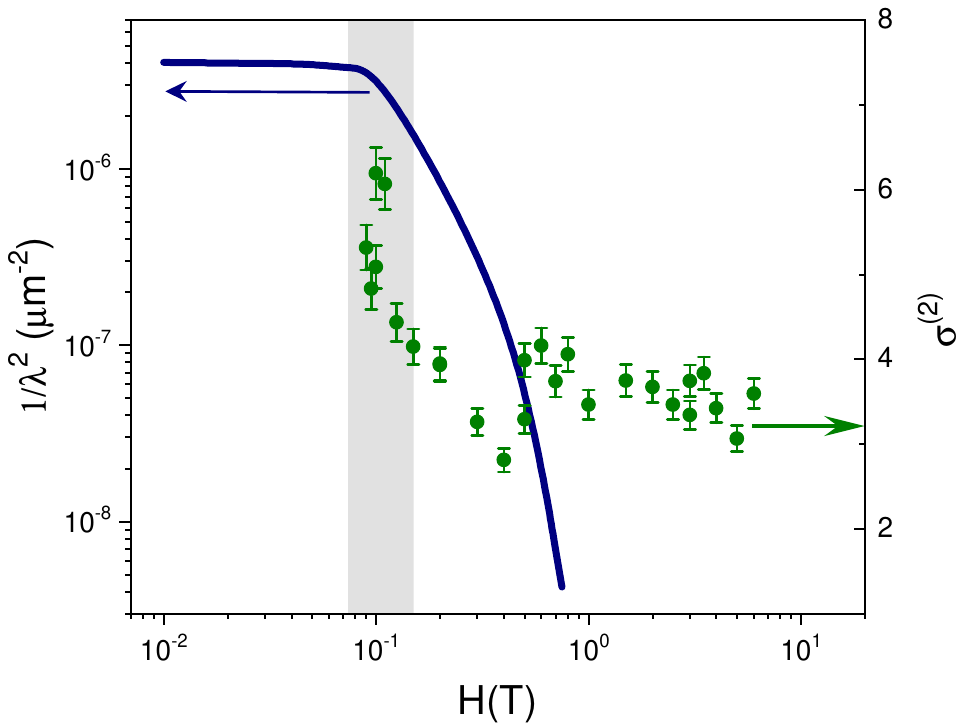}
		\small{\caption{Plots of $\lambda^{-2}$ (on left axis) and $\sigma^{(2)}$ (on right axis) versus $H$. Note that $\sigma^{(2)}$ drops sharply precisely where the super-fluid number density ($\propto \lambda^{-2}$) begins its sharp descent to zero.
				\label{fig:lambda-2}}}
	\end{center}
\end{figure}

Plots of $\sigma^{(2)}$ for both 2D- and 3D-SC  at different values of $R/R_N$ are shown in fig.~\ref{fig:SS}. $\sigma^{(2)}$ for 2D-SC  rises sharply close to zero-resistance state ($R/R_N << 0.1$), while it is close to 3 away from this regime. On the other hand, for the 3D-SC,  the value of $\sigma^{(2)}$  essentially fluctuates around 3  indicating that the  fluctuations are always Gaussian as expected for Ginzburg-Landau fluctuations in the case of a superconductor described by the  BCS theory. We plot $\sigma^{(2)}$ as a function of $R/R_N$ to emphasize that it diverges in the 2D-SC as $R \rightarrow0$. Recall that for both 2D- and 3D-SC, the noise $\mathcal{V}_{var}$ peaks near  $R/R_N \sim 0.5$. The appearance of non-Gaussian voltage-fluctuations in 2D-SC  over a regime in the phase-diagram which is far from that  where the variance of voltage-fluctuations peaks indicate that the non-Gaussianity is not due to the dynamical current-distribution, which is known to accompany the pinning-depinning of vortices~\cite{PhysRevLett.76.3049}. The fact that non-Gaussian voltage fluctuations are seen only in 2D-SC and not in 3D implies that its origin is distinct from the critical processes that cause voltage-fluctuations to peak in both 2D- and 3D-SC near the field-induced SC-normal transition. Instead, as established in our previous publications, the presence of a non-Gaussian component in the voltage fluctuation near the field-induced normal-SC transition is a consequence of long-range phase-fluctuations in 2D-SC~\cite{PhysRevLett.111.197001,PhysRevB.94.085104}. 

This conclusion finds support from the magnetic field dependence of the super-fluid number density  (which is proportional to $\lambda^{-2}$, $\lambda$ being the penetration depth). A plot of $\lambda^{-2}$ versus $H$ is shown in the left-axis of Fig.~\ref{fig:lambda-2}. On the right-axis $\sigma^{(2)}$ is plotted. One can see that at around the same value of magnetic field ($\sim 0.1~T$) where the super-fluid number density begins to crash towards zero with increasing $H$, the value of $\sigma^{(2)}$ also decreases sharply to the Gaussian value of 3. This  shows that the non-Gaussian voltage-fluctuations indeed arise due to the appearance of correlations in the system as one approaches the normal-SC transition.    Beyond this range of $H$, correlations between the vortices are lost, and the fluctuations are dominated by pinning-depinning of uncorrelated fluxoids.  

To conclude, in this article, we have explored the differences in vortex-dynamics in 2-dimensional and 3-dimensional superconductors. Our studies were performed on thin films of the same material system (NbN) prepared under identical conditions, thus eliminating the possibility of material-specific artifacts in comparison across dimensions, an issue that has plagued the community for a long time. We established that the observed fragility of 2D-SC to a perpendicular magnetic field as compared to the 3D-SC stems from the differing pinning-properties in these two dimensions. We show,  from two independent transport measurements carried out in two different regimes of vortex dynamics, that both the pinning-strength and the free-energy barrier to depinning are at least an order of magnitude stronger in 3D-SC than in 2D-SC. From voltage fluctuation measurements, we find that the dynamic process of flux pinning-depinning is similar in both 2D and 3D. Presence of non-Gaussian voltage fluctuations is observed only in the 2D-SC  and is understood to arise from long-range correlations between vortices close to the magnetic field-induced normal-superconducting transition.

A.B. acknowledges financial support from Nanomission, DST, Govt. of India  project SR/NM/NS-35/2012; SERB, DST, Govt. of India  and Indo-French Centre for the Promotion of Advanced Recearch (CEFIPRA). P.R. acknowledges funding from Department of Atomic Energy.  H.K.K. thanks CSIR, MHRD, Govt. of India for financial support. We acknowledge Vivas Bagwe for technical help in sample preparation.

%

\end{document}